\documentclass[12pt]{article}
\usepackage[margin=1in]{geometry}

\usepackage{graphicx}
\usepackage{url}
\usepackage{amssymb}
\usepackage{amsmath}
\usepackage{setspace}
\usepackage{authblk}
\usepackage{mathtools}
\usepackage{amsthm}
\usepackage{comment}
\usepackage{booktabs, tabularx, makecell, threeparttable, array, adjustbox}
\usepackage{enumitem}
\usepackage[round,authoryear]{natbib}
\usepackage{hyperref}

\newtheorem{theorem}{Theorem}[section]

\newtheorem{proposition}[theorem]{Proposition}

\newcolumntype{L}[1]{>{\raggedright\arraybackslash}p{#1}}
\newcolumntype{C}{>{\centering\arraybackslash\scriptsize}X}

\providecommand{\Pbb}{\mathbb{P}}
\providecommand{\E}{\mathbb{E}}
\providecommand{\Ber}{\mathrm{Bernoulli}}
\providecommand{\iid}{\overset{\mathrm{iid}}{\sim}}

\begin{document}
\doublespacing
\title{Likelihood-Ratio E-Value Monitoring for Benchmark-Based Decisions in Early-Phase Oncology Trials}
\author[1]{Masahiro Kojima}
\affil[1]{Department of Data Science for Business Innovation, Chuo University, 1-13-27 Kasuga, Bunkyo-ku, Tokyo 112-8551, Japan. E-mail: mkojima263@g.chuo-u.ac.jp}
\date{}

\maketitle

\noindent\textbf{Running title:} Calibrated e-value monitoring for phase II trials\\
\textbf{Corresponding author:} Masahiro Kojima, Department of Data Science for Business Innovation, Chuo University, 1-13-27 Kasuga, Bunkyo-ku, Tokyo 112-8551, Japan; E-mail: mkojima263@g.chuo-u.ac.jp.

\begin{abstract}\noindent
Early-phase oncology trials often require protocol-ready interim rules for deciding whether an experimental regimen shows sufficient activity relative to prespecified clinical benchmarks. Existing Bayesian optimal phase II designs provide calibrated count boundaries for such decisions, but evidence is quantified through posterior probabilities and sample-size-dependent cutoff functions. We propose calibrated e-value monitoring (CEVAM), a likelihood-ratio evidence framework that defines monitoring evidence directly relative to prespecified clinical benchmarks, without requiring a Bayesian analysis prior or sample-size-dependent posterior-probability cutoff functions. For a binary efficacy endpoint, CEVAM constructs efficacy and reverse-futility e-processes targeted to clinically meaningful benchmark rates and converts them into monotone response-count boundaries. We distinguish a fixed-threshold e-process version, which has an anytime-valid interpretation under optional monitoring, from planned-look calibrated versions designed for conventional finite-look phase II trials. In simulations based on binary benchmark settings used in existing posterior-cutoff phase II designs, the proposed tuned planned-look rule controlled the nominal type I error and achieved the smallest expected sample size across all evaluated settings while retaining similar rejection probabilities. In an application to an actual phase II breast cancer trial, CEVAM classified the reported pathological complete response result as sufficient evidence for early success stopping. Extensions to categorical and multicomponent endpoints are provided in the Supplementary Material. CEVAM offers an analysis-prior-free, protocol-ready likelihood-ratio evidence scale for binary benchmark monitoring in early-phase oncology trials.
\end{abstract}
\par\vspace{4mm}

\section{Introduction}
In early-phase oncology studies, investigators must decide whether an experimental regimen provides enough antitumor activity, acceptable toxicity, or both to justify continued development. These decisions are often made in small single-arm phase II studies, expansion cohorts after dose escalation, or dose-optimization programs in which several regimens are monitored against clinically meaningful benchmarks rather than compared formally with one another. Because patient populations may be limited and many participants have exhausted standard therapies, monitoring rules should protect patients from ineffective or unsafe regimens while allowing decisive favorable signals to be acted on without unnecessary additional accrual. In practice, such rules must also be simple enough to be specified before trial initiation as count boundaries for interim and final analyses.

The classical reference point for binary phase II monitoring is Simon's two-stage design \citep{Simon1989}, which provides exact type I and type II error control with early stopping for futility. Bayesian monitoring designs broaden this framework by allowing more flexible interim decision making, sequential posterior updating, and natural extensions to endpoints beyond a single binary response \citep{ThallSimon1994,Sambucini2008,CaiLiuYuan2014}. Among these designs, the Bayesian optimal phase II design (BOP2) is especially influential because it accommodates binary, ordinal, nested, co-primary, and efficacy--toxicity endpoints under a Dirichlet--multinomial framework, while still allowing explicit type I error calibration and trial-start tabulation of stopping boundaries \citep{ZhouLeeYuan2017}. BOP2 primarily provides posterior-probability futility stopping rules, and related extensions such as BOP2-FE incorporate early efficacy stopping in addition to futility monitoring, using calibrated posterior-probability cutoffs that vary with interim sample size \citep{XuHashimotoYimerTakeda2025}. These posterior-cutoff designs are practically attractive, but they also have features that motivate a complementary evidence-based formulation. First, even when the analysis prior is chosen to be vague, evidence is still quantified through posterior probabilities. Second, the operative stopping rule is expressed through sample-size-dependent posterior-probability cutoffs, or their futility--efficacy analogues, whose parametric forms are prespecified for computational and practical convenience and then calibrated to obtain desirable operating characteristics \citep{ZhouLeeYuan2017,XuHashimotoYimerTakeda2025}. Thus, the calibrated boundaries are highly useful for implementation, but the evidence scale itself is tied to posterior-probability thresholds. Third, because posterior threshold rules are not automatically valid under arbitrary optional stopping, their operating characteristics are fundamentally established by design-stage calibration rather than by a direct sequential validity argument.

Recent work on e-values and e-processes provides a useful alternative language for sequential evidence. An e-process is a nonnegative supermartingale under the null; by Ville's inequality, crossing a fixed threshold $1/\alpha$ yields time-uniform type I error control under optional monitoring \citep{GrunwaldDeHeideKoolen2024,RamdasGrunwaldVovkShafer2023,VovkWang2021}. This property is appealing in adaptive or flexible early-phase trials, where interim analyses may be irregular or additional reviews may occur. At the same time, many oncology phase II designs are implemented and compared at a fixed set of planned interim looks. We therefore distinguish the underlying fixed-threshold e-process construction from the planned-look calibrated implementation used for numerical comparison in this paper.

In this paper, we propose calibrated e-value monitoring (CEVAM), a likelihood-ratio evidence-monitoring framework for phase II oncology trials. CEVAM is tailored to the benchmark-monitoring problems addressed by BOP2 and BOP2-FE, but replaces posterior-probability evidence with likelihood-ratio evidence on an e-value scale. The key idea is to express clinically relevant endpoint information as Bernoulli projections and to evaluate evidence directly relative to prespecified clinical benchmarks. For a binary efficacy endpoint, this construction yields monotone response-count boundaries for early efficacy and futility decisions based on likelihood-ratio evidence and reverse likelihood-ratio evidence. For complex endpoints, the same projected-evidence principle can be applied to clinically meaningful components and combined according to the prespecified clinical decision structure. The novelty of CEVAM is not simply the use of e-values in phase II trials. Rather, CEVAM is designed to preserve the protocol-ready boundary tables and support for complex endpoint structures that make BOP2 and BOP2-FE attractive in practice, while avoiding the need for a Bayesian analysis prior. In contrast to posterior-cutoff designs, the evidence scale in CEVAM is defined by likelihood ratios relative to prespecified clinical benchmarks, so that the monitoring rule is built from an explicit evidence measure rather than from posterior-probability thresholds. This construction also distinguishes CEVAM from general e-value methodology for adaptive clinical trials, which has primarily emphasized anytime-valid monitoring, confidence sequences, betting-martingale constructions, and two-arm randomized settings. CEVAM instead targets a narrower but practically important oncology problem: protocol-ready, arm-level phase II benchmark monitoring with binary and complex endpoints. The main text develops the method for the binary efficacy endpoint setting, whereas extensions to ordinal or multinomial endpoints, nested efficacy endpoints, co-primary efficacy endpoints, and joint efficacy--toxicity endpoints are provided in the Supplementary Material.

This paper is organized as follows. Section~2 describes the proposed CEVAM methodology. Section~3 presents the simulation settings and results for the binary endpoint setting. Section~4 illustrates the proposed design using the TREND trial. Section~5 concludes with a discussion. Additional methodological details and simulation results for complex endpoints are provided in the Supplementary Material.

\section{Method}\label{sec:method}
In early oncology development, multiple candidate dose levels or regimens may be evaluated, sometimes using randomized allocation among dose arms. The monitoring rule considered in this paper is not intended to replace a formal between-dose comparison for final dose selection. Rather, it is applied at the individual dose-arm level to assess whether each candidate dose arm is sufficiently promising relative to prespecified clinical benchmarks. We therefore formulate CEVAM for a single dose arm. The same construction can be applied separately to each dose arm with its own response benchmarks, maximum sample size, and planned monitoring times.

We first present CEVAM for a binary efficacy endpoint, such as objective response, because this is the standard benchmark-monitoring setting in phase II oncology trials and yields simple response-count boundaries for early efficacy and futility decisions. Extensions to ordinal or multinomial endpoints, nested efficacy endpoints, co-primary efficacy endpoints, and joint efficacy--toxicity endpoints are developed in the Supplementary Material using the same projected-evidence principle.

\subsection{Binary e-processes for efficacy and futility}\label{sec:binary-lr}
Let $X_1,X_2,\ldots$ denote independent patient-level binary outcomes, where $X_i=1$ indicates treatment response and $X_i=0$ indicates nonresponse. We write $X_i \iid \Ber(p)$, with $p\in[0,1]$, and let $R_n=\sum_{i=1}^n X_i$ denote the cumulative number of responses among the first $n$ patients. The design specifies a maximum sample size $N$ and a set of planned interim looks $\mathcal N\subseteq\{1,\ldots,N\}$. Let $p_0$ denote the unacceptable response rate and let $p_1>p_0$ denote a clinically meaningful design alternative. Throughout the likelihood-ratio construction, we assume $0<p_0<p_1<1$.

The parameter $\alpha$ is the prespecified evidence-threshold parameter for efficacy in the fixed-threshold e-process version. Because the efficacy stopping threshold is $1/\alpha$, $\alpha$ also has the usual type I error interpretation for the efficacy claim in the anytime-valid fixed-threshold version. When futility monitoring is used, $\beta$ is the prespecified evidence-threshold parameter for the reverse-evidence futility rule. It determines the futility threshold $1/\beta$; larger values of $\beta$ make futility stopping more aggressive. This $\beta$ is not the type I error level for the efficacy claim, but in the fixed-threshold reverse e-process rule it bounds the probability of incorrectly stopping for futility under $H_0^{\mathrm{fut}}:p\ge p_1$.

The efficacy claim is formulated as the one-sided benchmark hypothesis
\[
H_0^{\mathrm{eff}}:p\le p_0
\qquad\text{versus}\qquad
H_1^{\mathrm{eff}}:p>p_0 .
\]
To construct an e-process against this composite null, we prespecify the design alternative $p_1$. The value $p_1$ is not intended to exhaust the alternative space $p>p_0$; rather, it specifies the betting alternative at which the likelihood-ratio evidence process is targeted, analogous to the design alternative used for power calculations in conventional phase II designs. Specifically, we compare the simple design alternative $p=p_1$ with the least favorable boundary point $p=p_0$ of the composite null and define
\begin{align}
E_n^{\mathrm{LR}}(p_1,p_0)
&=
\prod_{i=1}^n
\frac{p_1^{X_i}(1-p_1)^{1-X_i}}
     {p_0^{X_i}(1-p_0)^{1-X_i}} \notag\\
&=
\left(\frac{p_1}{p_0}\right)^{R_n}
\left(\frac{1-p_1}{1-p_0}\right)^{n-R_n}.
\label{eq:lr}
\end{align}
Thus, $E_n^{\mathrm{LR}}(p_1,p_0)$ is a likelihood-ratio evidence process targeted to the clinically meaningful response rate $p_1$, while its validity is with respect to the full composite null $p\le p_0$. Rejection therefore indicates sufficient evidence against $p\le p_0$, not evidence that the true response rate is exactly $p_1$.

For futility, we use the reverse comparison. A direct way to formalize lack of promise is
\begin{align}
H_0^{\mathrm{fut}}:p\ge p_1
\qquad\text{versus}\qquad
H_1^{\mathrm{fut}}:p<p_1 .
\end{align}
Here $p_1$ is the clinically meaningful target response rate, and $p_0$ serves as a representative nonpromising response rate. Define the reverse likelihood-ratio process
\begin{align}
F_n^{\mathrm{LR}}(p_0,p_1)
&=
\prod_{i=1}^n
\frac{p_0^{X_i}(1-p_0)^{1-X_i}}
     {p_1^{X_i}(1-p_1)^{1-X_i}} \notag\\
&=
\left(\frac{p_0}{p_1}\right)^{R_n}
\left(\frac{1-p_0}{1-p_1}\right)^{n-R_n}.
\label{eq:rev-lr}
\end{align}
This statistic measures how much more the accumulated data support the representative nonpromising rate $p_0$ than the clinically meaningful target $p_1$. Therefore, crossing a futility threshold does not mean that the observed response rate is simply below $p_1$. Rather, it means that the likelihood-ratio evidence in favor of $p_0$ over $p_1$ is sufficiently large.

\begin{theorem}\label{thm:eff-valid}
Assume $X_1,X_2,\ldots$ are independent Bernoulli random variables with common success probability $p$, and let $0<p_0<p_1<1$. For any $0<\alpha<1$ and $0<\beta<1$, the following statements hold. The efficacy process $\{E_n^{\mathrm{LR}}(p_1,p_0):n\ge0\}$ is a nonnegative supermartingale under every distribution in the composite null $H_0^{\mathrm{eff}}:p\le p_0$. Consequently,
\begin{align}
\sup_{p\le p_0}
\Pbb_p\!\left(
\sup_{n\ge1}E_n^{\mathrm{LR}}(p_1,p_0)\ge \frac{1}{\alpha}
\right)
\le \alpha.
\end{align}
Similarly, the reverse process $\{F_n^{\mathrm{LR}}(p_0,p_1):n\ge0\}$ is a nonnegative supermartingale under every distribution in the composite null $H_0^{\mathrm{fut}}:p\ge p_1$. Consequently,
\begin{align}
\sup_{p\ge p_1}
\Pbb_p\!\left(
\sup_{n\ge1}F_n^{\mathrm{LR}}(p_0,p_1)\ge \frac{1}{\beta}
\right)
\le \beta.
\end{align}
\end{theorem}
\noindent
The proof of Theorem~\ref{thm:eff-valid} is provided in the Supplementary Material.

The corresponding fixed-threshold binary monitoring rules are
\begin{align*}
\tau_{\mathrm{eff}}
&=
\inf\left\{
n\in\mathcal N:
E_n^{\mathrm{LR}}(p_1,p_0)\ge \frac{1}{\alpha}
\right\},\\
\tau_{\mathrm{fut}}
&=
\inf\left\{
n\in\mathcal N:
F_n^{\mathrm{LR}}(p_0,p_1)\ge \frac{1}{\beta}
\right\}.
\end{align*}
Thus, once $(p_0,p_1,\alpha,\beta,N,\mathcal N)$ are prespecified, the fixed-threshold binary design is fully determined. At each planned look, the observed likelihood-ratio evidence is simply compared with the fixed thresholds $1/\alpha$ for efficacy and $1/\beta$ for futility. The two threshold parameters play different roles and need not be equal: $\alpha$ controls the efficacy evidence threshold, whereas $\beta$ controls the aggressiveness of the reverse-evidence futility rule. Larger values of $\beta$ correspond to lower futility evidence thresholds and therefore more aggressive futility stopping. Because these thresholds are fixed in advance, this version does not require grid-search calibration of sample-size-dependent posterior-probability cutoff functions.

A key practical feature of BOP2 and BOP2-FE is that stopping boundaries can be written as a table of counts before the trial starts. The likelihood-ratio e-process construction preserves this property.

\begin{proposition}[Monotonicity and pretabulated binary boundaries]
\label{prop:binary-boundaries}
Let $0<p_0<p_1<1$. For fixed $n$ and $\alpha,\beta\in(0,1)$,
$E_n^{\mathrm{LR}}(p_1,p_0)$ is strictly increasing in $R_n$, whereas
$F_n^{\mathrm{LR}}(p_0,p_1)$ is strictly decreasing in $R_n$. Hence there exist
integer boundaries $r_{\mathrm{eff}}(n;\alpha)$ and $r_{\mathrm{fut}}(n;\beta)$,
possibly outside $\{0,\ldots,n\}$, such that
\begin{align}
E_n^{\mathrm{LR}}(p_1,p_0)\ge \frac{1}{\alpha}
\qquad\Longleftrightarrow\qquad
R_n\ge r_{\mathrm{eff}}(n;\alpha),
\end{align}
and
\begin{align}
F_n^{\mathrm{LR}}(p_0,p_1)\ge \frac{1}{\beta}
\qquad\Longleftrightarrow\qquad
R_n\le r_{\mathrm{fut}}(n;\beta).
\end{align}
Explicitly,
\begin{align}
r_{\mathrm{eff}}(n;\alpha)
=
\left\lceil
\frac{\log(1/\alpha)-n\log\big((1-p_1)/(1-p_0)\big)}
{\log\big(p_1(1-p_0)/(p_0(1-p_1))\big)}
\right\rceil,
\end{align}
and
\begin{align}
r_{\mathrm{fut}}(n;\beta)
=
\left\lfloor
\frac{\log(1/\beta)-n\log\big((1-p_0)/(1-p_1)\big)}
{\log\big(p_0(1-p_1)/(p_1(1-p_0))\big)}
\right\rfloor.
\end{align}
\end{proposition}
\noindent
The proof of Proposition~\ref{prop:binary-boundaries} is provided in the Supplementary Material.

At look $n$, the continuation region is therefore
\begin{align}
r_{\mathrm{fut}}(n;\beta) < R_n < r_{\mathrm{eff}}(n;\alpha).
\end{align}
If this interval is empty at some early look, then the pair $(\alpha,\beta)$ is too aggressive for the chosen $(p_0,p_1,N,\mathcal N)$. In that case, one may reduce $\beta$, reduce $\alpha$, or postpone the first interim look. This is not a conceptual defect of the no-tuning design; it reflects the fact that very low evidence thresholds over a short horizon can make early efficacy and futility regions overlap.

\subsection{Planned-look calibration of efficacy and futility thresholds}
\label{subsec:planned-look-calibration}
The fixed-threshold e-process version is simple and has a direct sequential-validity interpretation. In this version, $\alpha$ and $\beta$ are prespecified evidence-threshold parameters, and the monitoring rules compare the observed likelihood-ratio evidence with $1/\alpha$ and $1/\beta$, respectively. This version is fully determined once $(p_0,p_1,\alpha,\beta,N,\mathcal N)$ are fixed and retains the direct optional-monitoring interpretation described in Theorem~\ref{thm:eff-valid}.

For conventional phase II trials monitored only at prespecified interim looks, however, the fixed-threshold rule can be conservative at the planned looks. We therefore introduce planned-look calibrated CEVAM rules for such finite-look designs. These rules remain likelihood-ratio evidence rules, but their threshold parameters are selected at the design stage to obtain desirable operating characteristics for the prespecified monitoring schedule. This calibration is analogous in purpose to the grid-search calibration used in BOP2 and BOP2-FE, but the calibrated quantities are evidence-threshold parameters rather than posterior-probability cutoff-function parameters.

Let $\alpha^\ast$ denote the nominal type I error level for the efficacy claim. In the fixed-threshold anytime-valid version, the efficacy threshold is set directly by $\alpha^\ast$. In the planned-look calibrated implementation, however, we distinguish the nominal target $\alpha^\ast$ from the efficacy evidence-threshold parameter $\alpha_{\mathrm{eff}}$. Let $\mathcal A$ be a prespecified grid of candidate values for $\alpha_{\mathrm{eff}}$ and let $\mathcal B$ be a prespecified grid of candidate futility-threshold parameters. For each candidate $(\alpha_{\mathrm{eff}},\beta)$, construct the count boundaries $r_{\mathrm{eff}}(n;\alpha_{\mathrm{eff}})$ and $r_{\mathrm{fut}}(n;\beta)$ using Proposition~\ref{prop:binary-boundaries}, and evaluate the planned-look operating characteristics under the design null $p_0$ and design alternative $p_1$:
\begin{equation}
\begin{aligned}
\mathrm{TypeI}(\alpha_{\mathrm{eff}},\beta)
&=
\Pbb_{p_0}\{\text{reject }H_0^{\mathrm{eff}}\},\\
\mathrm{Pow}(\alpha_{\mathrm{eff}},\beta)
&=
\Pbb_{p_1}\{\text{reject }H_0^{\mathrm{eff}}\},\\
\mathrm{EN}_0(\alpha_{\mathrm{eff}},\beta)
&=
\E_{p_0}(N_{\mathrm{used}}).
\end{aligned}
\label{eq:calib-ocs}
\end{equation}
Here, the three quantities in \eqref{eq:calib-ocs} denote, respectively,
the planned-look type I error under the design null, the rejection
probability under the design alternative, and the expected sample size
under the design null, where \(N_{\mathrm{used}}\) denotes the sample size
at the time of stopping or at the final analysis if the trial is not
stopped early.

We consider two planned-look calibrated implementations. The first implementation fixes the reverse-evidence futility parameter at a prespecified value, rather than selecting it by calibration. We refer to this as the no-tuning futility implementation, abbreviated as CEVAM-NT. In CEVAM-NT, the reverse-evidence futility parameter is fixed at $\beta_0$, and only the efficacy-threshold parameter $\alpha_{\mathrm{eff}}$ is calibrated over $\mathcal A$. Thus, CEVAM-NT is ``no-tuning'' with respect to the futility parameter, but it is still planned-look calibrated for efficacy.

The second implementation tunes both the efficacy and futility evidence thresholds. We refer to this tuned implementation as CEVAM-T. In CEVAM-T, both $\alpha_{\mathrm{eff}}$ and $\beta$ are selected over $\mathcal A\times\mathcal B$. Candidate designs are required to control the planned-look type I error at the nominal level $\alpha^\ast$, up to a prespecified small tolerance used to accommodate discreteness of count boundaries. Among feasible candidates, the implementation first selects the design whose planned-look type I error is closest to $\alpha^\ast$ without exceeding the allowed tolerance, then breaks ties by larger rejection probability under the design alternative, smaller expected sample size under the design null, and larger early-termination probability under the design null.

This calibration step changes the interpretation of the design. The fixed-threshold version with efficacy threshold parameter $\alpha^\ast$ retains the direct anytime-valid e-process guarantee. By contrast, the planned-look calibrated versions, CEVAM-NT and CEVAM-T, are calibrated for the prespecified monitoring schedule, and their type I error control is established by design-stage operating-characteristic calculations. Therefore, these calibrated versions should not be described as providing arbitrary optional-monitoring validity. Despite this calibration step, the CEVAM implementation remains simple in three respects. The monitoring statistic is a likelihood-ratio evidence process and does not require a beta or Dirichlet analysis prior. The resulting efficacy and futility rules are monotone in event counts, so they can be summarized as protocol-ready boundary tables. Moreover, the calibrated parameters retain evidence-scale interpretations: $\alpha_{\mathrm{eff}}$ controls the planned-look efficacy threshold and $\beta$ controls the reverse-evidence futility threshold.

For complex endpoints, the same idea can be applied to clinically meaningful Bernoulli projections and combined according to the prespecified clinical OR/AND decision rule. Because the main text focuses on the binary endpoint setting, the full construction and simulation results for nested efficacy, co-primary efficacy, and joint efficacy--toxicity endpoints are provided in the Supplementary Material.

The behavior of the likelihood-ratio evidence process can be understood through its expected log-growth. For a binary endpoint, the per-patient expected log-growth of the efficacy evidence under response rate $p$ is
\begin{equation}
g(p;p_1,p_0)
=
p\log\!\left(\frac{p_1}{p_0}\right)
+
(1-p)\log\!\left(\frac{1-p_1}{1-p_0}\right).
\label{eq:growth}
\end{equation}
When this quantity is large and positive, efficacy evidence accumulates rapidly and early efficacy stopping is more likely. The reverse likelihood-ratio evidence has an analogous expected log-growth for futility. Thus, CEVAM is expected to reduce the sample size primarily when the data-generating distribution is clearly promising or clearly nonpromising relative to the prespecified benchmarks. Further discussion of these log-growth approximations for expected stopping times and sample-size reduction is provided in the Supplementary Material.

\section{Simulation study}
We conducted Monte Carlo simulation studies to evaluate the operating characteristics of the proposed CEVAM design.

\subsection{Simulation configuration}

The simulation study evaluated the binary efficacy endpoint setting, which corresponds to the one-dimensional likelihood-ratio construction developed in the main text. The simulation settings were chosen with reference to the binary endpoint examples in BOP2 \citep{ZhouLeeYuan2017} and BOP2-FE \citep{XuHashimotoYimerTakeda2025}, so that the proposed CEVAM design could be compared with existing protocol-ready phase II monitoring designs under clinically relevant benchmark configurations.

For the binary efficacy endpoint setting, the maximum sample size was set to $N=40$, with planned interim looks at $\mathcal N=\{10,15,20,25,30,35,40\}$. The design null and design alternative were specified as $p_0=0.20$ and $p_1=0.40$, respectively, following the corresponding binary benchmark setting used in the BOP2-FE simulation examples. The nominal type I error level for the efficacy claim was set to $\alpha^\ast=0.10$.

For CEVAM-NT, the reverse-evidence futility parameter was fixed at the prespecified value $\beta_0=0.10$, and the efficacy evidence-threshold parameter $\alpha_{\mathrm{eff}}$ was selected by grid search over $\mathcal A=\{0.01,0.02,\ldots,1.00\}$. For CEVAM-T, both the efficacy evidence-threshold parameter $\alpha_{\mathrm{eff}}$ and the futility-threshold parameter $\beta$ were selected by grid search over $\mathcal A=\{0.01,0.02,\ldots,1.00\}$ and $\mathcal B=\{0.01,0.02,\ldots,1.00\}$. Candidate designs were required to control the planned-look type I error at the nominal level $\alpha^\ast$, allowing only the prespecified discreteness tolerance used in the calibration. Among feasible candidates, the final design was selected according to the calibration criterion described in Section~\ref{subsec:planned-look-calibration}.

The numerical study compared five protocol-ready designs:
\begin{enumerate}[label=(\arabic*),leftmargin=1.5em]
\item BOP2-F, the original futility-only BOP2 design \citep{ZhouLeeYuan2017};
\item BOP2-FE-1, the BOP2-FE design with power-function futility and efficacy cutoffs \citep{XuHashimotoYimerTakeda2025};
\item BOP2-FE-2, the BOP2-FE design with a power-function futility cutoff and an O'Brien--Fleming-type efficacy cutoff \citep{XuHashimotoYimerTakeda2025};
\item CEVAM-NT, the planned-look calibrated e-value monitoring design with calibrated efficacy threshold and fixed reverse-evidence futility threshold;
\item CEVAM-T, the planned-look calibrated e-value monitoring design with calibrated efficacy and futility thresholds.
\end{enumerate}

The primary operating characteristics were the probability of rejecting the null hypothesis (PRN), early termination for futility (ETF), early termination for efficacy (ETE), probability of early termination (PET), and expected sample size (EN). The comparison was intentionally focused on planned-look calibrated rules because BOP2-F and BOP2-FE are also evaluated as pretabulated planned-look designs.

In addition to the usual operating characteristics, we descriptively summarized how each design allocated false-positive probability over the planned looks. This analysis was performed only under the design null hypothesis and was not used to construct the stopping boundaries. For design $m$ and planned looks $n_1<\cdots<n_K$, define
\begin{equation}
 A_m(k)
 =
 \Pbb_{H_0}\{\text{the trial rejects }H_0\text{ for efficacy at or before look }k\},
 \qquad k=1,\ldots,K.
\label{eq:alpha-spending-main}
\end{equation}
The increment
\begin{equation}
\Delta_m(k)=A_m(k)-A_m(k-1),\qquad A_m(0)=0,
\end{equation}
represents the additional false-positive probability allocated at look $k$. Although BOP2-type designs are not derived from the classical group-sequential alpha-spending framework, the curve $A_m(k)$ provides a useful descriptive analogue of an alpha-spending curve. In the present comparison, $A_m(k)$ was estimated by simulating trials under the binary design null $p=p_0$ only; details are given in the Supplementary Material.

Supplementary analyses were also conducted to evaluate settings beyond the binary efficacy endpoint. Specifically, the Supplementary Material reports additional simulations for nested efficacy endpoints, co-primary efficacy endpoints, and joint efficacy--toxicity monitoring, using componentwise projected CEVAM rules combined according to the corresponding clinical OR/AND decision structures. The Supplementary Material also reports optional-monitoring simulations under unplanned looks and optional continuation, which were conducted to evaluate the fixed-threshold anytime-valid CEVAM rule in the setting for which it is primarily intended and to contrast it with planned-look calibrated boundary-table designs. These supplementary analyses are intended to illustrate the extension of CEVAM to complex endpoint structures and the distinction between fixed-threshold e-process monitoring and planned-look calibrated boundary tables, rather than to redefine the primary comparison in the main text.

\subsection{Simulation results}
The stopping and efficacy boundaries used in the simulation are shown in Table~\ref{tab:phase2-boundary-example1-pair1}, and the corresponding simulation results are presented in Table~\ref{tab:phase2-sim-example1-pair1}. Under the design null hypothesis with a true efficacy probability of 0.20, the proposed CEVAM designs, as well as the comparator designs, maintained the nominal type I error level of 0.10. Among the evaluated designs, CEVAM-T achieved the smallest expected sample size under the null setting.

As the true efficacy probability increased, CEVAM-T consistently achieved the smallest expected sample size among the evaluated designs while maintaining rejection probabilities close to those of BOP2-FE. This pattern shows that the tuned likelihood-ratio evidence rule can improve trial efficiency across the evaluated binary benchmark settings without materially compromising the probability of detecting a promising regimen. Although the rejection probability of CEVAM-T was slightly lower than that of BOP2-FE in some alternative settings, the reduction in expected sample size was substantial, supporting the use of the tuned CEVAM rule as an efficient planned-look monitoring design in this binary endpoint setting.

\begin{figure}[htbp]
\centering
\includegraphics[width=0.70\textwidth]{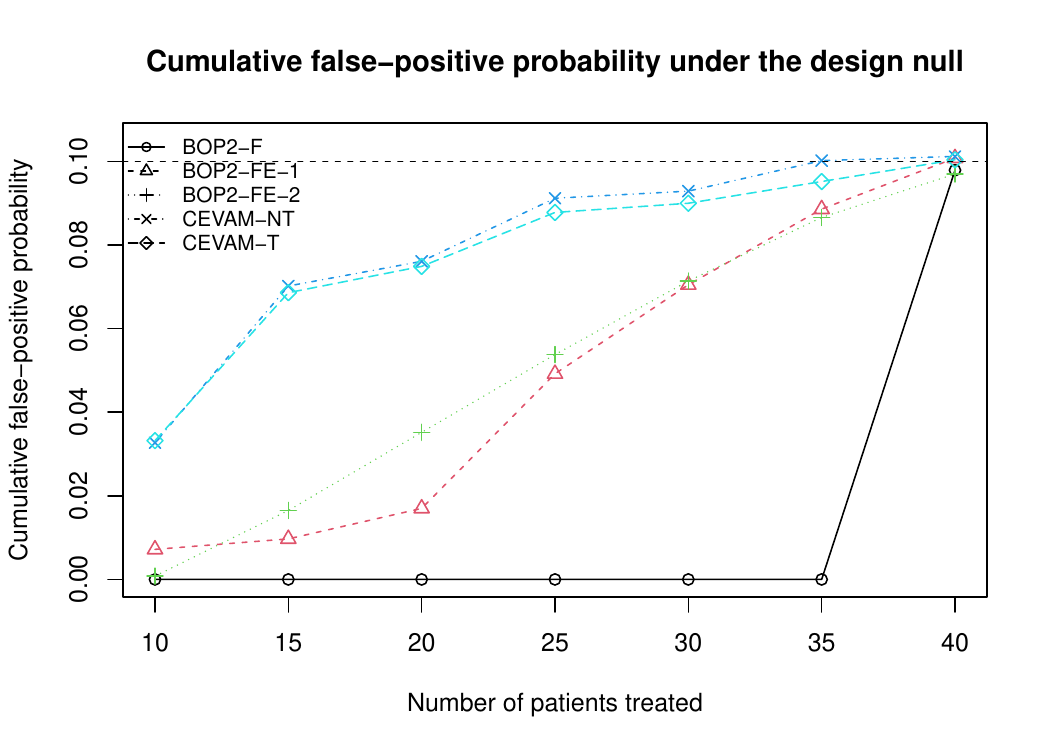}
\caption{Descriptive cumulative false-positive probability curves under the binary design null hypothesis. The vertical axis shows the estimated cumulative false-positive probability, $\widehat A_m(k)$, by planned look. These curves are descriptive analogues of alpha-spending curves and are not used to construct the stopping boundaries.}
\label{fig:alpha-spending-curves}

\noindent\textbf{Alt text:} Line plot showing cumulative false-positive probability by planned interim look for five phase II monitoring designs under the binary design null. The CEVAM-NT and CEVAM-T curves allocate more false-positive probability at earlier looks than the comparator designs, whereas all designs approach the nominal type I error level by the final analysis.
\end{figure}

Figure~\ref{fig:alpha-spending-curves} displays the resulting cumulative false-positive probability curves. The proposed CEVAM designs tended to spend the type I error probability earlier than the comparator designs. In particular, CEVAM-NT and CEVAM-T allocated a larger fraction of the total false-positive probability at the early interim looks, whereas BOP2-F spent almost no type I error probability until the final look and BOP2-FE showed a more gradual accumulation. By the final analysis, all designs were close to the nominal type I error level.

\section{Application to the TREND trial}\label{sec:application}
We illustrate the proposed binary CEVAM design using the published TREND trial~\cite{zhang2025efficacy}. This was a prospective single-arm phase II study of neoadjuvant tislelizumab plus chemotherapy for triple-negative breast cancer. The primary endpoint was pathological complete response (pCR), defined as ypT0/Tis ypN0. The protocol specified a maximum sample size of 65 patients and one interim efficacy and safety analysis after enrollment of 48 patients. Under the original Bayesian optimal phase II monitoring rule, the trial would be terminated for futility if 20 or fewer patients achieved pCR, whereas accrual could be terminated early for success if at least 30 patients achieved pCR. The null and target pCR rates were 0.41 and 0.56, respectively. This design was reported to have a one-sided type I error rate of 0.05 and 80.0\% power when the true pCR rate was 0.56.

The final publication reported that 53 patients were enrolled, 44 patients were included in the efficacy-evaluable set, and 30 of 44 patients achieved pCR. The report also states that the trial was terminated early because the primary endpoint was met. Because the exact aggregate pCR count at the protocol-specified 48-patient interim analysis was not reported, we used the published efficacy-evaluable result, \(x=30\) pCRs among \(n=44\) patients, as a retrospective re-analysis point. This application is therefore intended to illustrate how CEVAM would classify the reported early-termination result under the same benchmark rates and maximum sample size, rather than to reconstruct the original trial conduct.

Using \(p_0=0.41\), \(p_1=0.56\), \(\alpha^\ast=0.05\), and \(N=65\), we constructed binary CEVAM boundaries at \(n=44\) and \(n=65\). Table~\ref{tab:trend-cevam-boundaries} shows two planned-look CEVAM implementations, using the same notation as in the simulation studies. CEVAM-NT fixes the reverse-evidence futility parameter at \(\beta=0.20\) and calibrates the efficacy threshold, whereas CEVAM-T jointly calibrates the efficacy and futility evidence thresholds over the specified analysis looks.

At the retrospective re-analysis point, the published result was \(x=30\) pCRs among \(n=44\) efficacy-evaluable patients. Under CEVAM-NT, the efficacy boundary at \(n=44\) was 25 responses, and under CEVAM-T it was 24 responses. The reported result exceeded both efficacy boundaries. Therefore, both CEVAM implementations would have classified the reported aggregate result as sufficient evidence of efficacy and would have supported early termination for success.

The tuned CEVAM-T rule produced more permissive futility boundaries than CEVAM-NT at both analysis looks. At \(n=44\), the futility boundary was 21 for CEVAM-T and 18 for CEVAM-NT; at \(n=65\), the corresponding boundaries were 31 and 28. The efficacy boundary was also slightly lower for CEVAM-T at the retrospective look, 24 versus 25 responses, whereas both implementations had the same final efficacy boundary of 35 responses at \(n=65\). This behavior is consistent with the role of CEVAM-T in the simulation studies: CEVAM-NT retains a simpler implementation by fixing the reverse-evidence futility parameter, whereas CEVAM-T further tunes both efficacy and futility thresholds for the specified finite set of looks.

This application illustrates how the proposed CEVAM rules can be applied using the benchmark rates and maximum sample size reported for a published phase II oncology trial. Although the original TREND protocol planned its interim analysis after enrollment of 48 patients and the published report provides the aggregate efficacy-evaluable result at \(n=44\), the observed pCR count of 30 exceeded both CEVAM efficacy boundaries at this retrospective re-analysis point. Thus, under the same benchmark rates and maximum sample size, the proposed likelihood-ratio evidence monitoring rules would have supported early success stopping for the reported TREND trial result.

\section{Discussion}
We proposed CEVAM, a calibrated e-value monitoring framework for benchmark-based phase II oncology trials. In the binary efficacy endpoint setting, CEVAM uses likelihood-ratio evidence and reverse likelihood-ratio evidence to construct simple response-count boundaries for early efficacy and futility monitoring. The method was motivated by the practical strengths of BOP2 and BOP2-FE, including explicit operating-characteristic calibration, early stopping, and protocol-ready boundary tables. CEVAM preserves this implementation format but changes the evidence scale: monitoring evidence is defined directly relative to prespecified clinical benchmarks through likelihood ratios, without requiring a Bayesian analysis prior or sample-size-dependent posterior-probability cutoff functions.

The main contribution of CEVAM is to connect likelihood-ratio e-processes with pretabulated count-boundary designs for early-phase oncology trials. The efficacy likelihood-ratio process is valid under the composite null \(p\le p_0\), and the reverse likelihood-ratio process is valid for the futility claim under \(p\ge p_1\). Because both processes are monotone functions of the cumulative response count, the resulting rules can be expressed as integer futility and efficacy boundaries at each interim look. This property is essential for clinical protocol implementation and makes the design directly comparable with established BOP2-type boundary-table designs.

We distinguished the fixed-threshold e-process construction from the planned-look calibrated implementations used for comparison in conventional finite-look trials. The fixed-threshold version provides the direct anytime-valid implementation of the likelihood-ratio e-process rule and is suited to settings in which additional reviews, irregular monitoring, or optional continuation may occur. By contrast, CEVAM-NT and CEVAM-T are planned-look calibrated versions intended for prespecified monitoring schedules. CEVAM-NT fixes the reverse-evidence futility parameter and calibrates the efficacy threshold, whereas CEVAM-T tunes both efficacy and futility thresholds over the specified finite set of looks. Thus, the CEVAM framework provides two practically distinct implementations: an anytime-valid fixed-threshold rule for flexible monitoring and calibrated finite-look boundary-table rules for conventional phase II trial designs.

In the binary endpoint simulation, CEVAM-T achieved the smallest expected sample size among all evaluated designs across the full set of true response probabilities while maintaining the planned-look type I error at the nominal level. These efficiency gains were obtained with rejection probabilities close to those of BOP2-FE, indicating that the likelihood-ratio evidence scale can improve trial efficiency without materially compromising power. This behavior reflects the design objective of CEVAM-T: the futility threshold is tuned together with the efficacy threshold so that decisive evidence can lead to earlier stopping under the prespecified monitoring schedule. The results therefore support CEVAM-T as an efficient likelihood-ratio evidence alternative to posterior-cutoff monitoring, particularly when investigators value interpretable evidence thresholds and protocol-ready count boundaries.

The TREND trial application illustrates how CEVAM can be applied to a published binary oncology endpoint. Using the reported pCR benchmarks and maximum sample size, CEVAM-NT and CEVAM-T classified the reported aggregate result of 30 pCRs among 44 efficacy-evaluable patients as sufficient evidence for early success stopping. This application does not reconstruct the exact conduct of the original trial, because the published report did not provide the exact pCR count at the protocol-specified 48-patient interim analysis. Rather, it demonstrates that the proposed likelihood-ratio evidence rules can be specified from clinically reported design parameters and can support an early-success decision using the reported aggregate pCR result.

Several extensions are natural. First, the main text focused on immediate binary endpoint monitoring. Delayed or pending outcomes, which are common in oncology trials, could be incorporated by combining the proposed evidence process with prediction, weighting, or imputation methods for incomplete response information. Second, the planned-look calibrated CEVAM rules are designed for the monitoring schedule used during calibration, whereas the fixed-threshold e-process version provides the corresponding implementation when additional or irregular reviews are anticipated. Third, the Supplementary Material shows that the same projected-evidence principle can be extended to nested efficacy, co-primary efficacy, and joint efficacy--toxicity endpoints. These extensions preserve the protocol-ready componentwise boundary format used in BOP2-type designs and provide a foundation for further development in multicomponent phase II monitoring.

In conclusion, CEVAM provides an analysis-prior-free, protocol-ready likelihood-ratio evidence framework for benchmark-based monitoring in early-phase oncology trials. By converting likelihood-ratio e-processes into monotone response-count boundaries, CEVAM retains the practical boundary-table format required for clinical protocols while replacing posterior-probability cutoffs with interpretable evidence thresholds defined relative to prespecified clinical benchmarks. In the binary benchmark setting, the tuned planned-look implementation, CEVAM-T, improved trial efficiency by reducing expected sample size across all evaluated scenarios while controlling type I error at the nominal level and preserving rejection probabilities close to those of established BOP2-FE designs. The fixed-threshold e-process version further extends the framework to settings in which unplanned reviews, irregular monitoring, or optional continuation are anticipated. These features make CEVAM a flexible and practically implementable likelihood-ratio evidence alternative to posterior-cutoff monitoring for early-phase oncology trials. Future work should address delayed outcomes, threshold calibration under alternative look schedules, and integration with dose-optimization and multi-arm early-phase trial designs.

\section*{Acknowledgments}
This work was supported by JSPS KAKENHI (Grant Number JP26K21185). ChatGPT was used only to assist with English language proofreading. The author reviewed and verified the final manuscript and takes full responsibility for its content.

\section*{Data availability}
No individual patient-level data were analyzed in this methodological study. The TREND trial application used only aggregate information reported in the published article and protocol. The R code used to generate the CEVAM boundary tables, simulation results, and TREND trial re-analysis is available at \url{https://github.com/masahikoji/cevam-phase2-monitoring}.

\newpage
\bibliographystyle{plainnat}
\bibliography{main}
\newpage

\section*{Supplementary Material}
Web Appendices, Web Tables, Web Figures, and R code referenced in Sections~2--5 are available with this paper at the Biometrics website on Oxford Academic. The R code used to generate the boundary tables, simulation results, alpha-spending plots, and TREND trial re-analysis is also available at the public GitHub repository listed in the Data Availability statement.


\begin{table}[htbp]
\centering
\begin{threeparttable}
\caption{Stopping and efficacy boundaries}
\label{tab:phase2-boundary-example1-pair1}
\begin{tabular}{llccccccc}
\toprule
Method & Rule & 10 & 15 & 20 & 25 & 30 & 35 & 40 \\
\midrule
BOP2-F & Futility: \# eff $\leq$ & 1 & 2 & 4 & 5 & 7 & 9 & 10 \\
\addlinespace
BOP2-FE-1 & Futility: \# eff $\leq$ & 1 & 2 & 3 & 5 & 7 & 8 & 11 \\
 & Efficacy: \# eff $\geq$ & 6 & 8 & 9 & 9 & 10 & 11 & 12 \\
\addlinespace
BOP2-FE-2 & Futility: \# eff $\leq$ & 1 & 2 & 4 & 5 & 7 & 8 & 11 \\
 & Efficacy: \# eff $\geq$ & 7 & 7 & 8 & 9 & 10 & 11 & 12 \\
\addlinespace
CEVAM-NT & Futility: \# eff $\leq$ & 0 & 2 & 3 & 4 & 6 & 7 & 9 \\
 & Efficacy: \# eff $\geq$ & 5 & 6 & 8 & 9 & 11 & 12 & 14 \\
\addlinespace
CEVAM-T & Futility: \# eff $\leq$ & 1 & 3 & 4 & 5 & 7 & 8 & 10 \\
 & Efficacy: \# eff $\geq$ & 5 & 6 & 8 & 9 & 11 & 12 & 13 \\
\bottomrule
\end{tabular}
\begin{tablenotes}[flushleft]
\footnotesize \item ORR, objective response rate. Pair 1 corresponds to the BOP2-FE Table 1 binary setting.
\end{tablenotes}
\end{threeparttable}
\end{table}

\begin{table}[htbp]
\centering
\begin{threeparttable}
\caption{Operating characteristics for Scenario 1, pair 1.}
\label{tab:phase2-sim-example1-pair1}
\begin{tabular}{lcccccc}
\toprule
True eff prob & Method & PRN (\%) & ETF (\%) & ETE (\%) & PET (\%) & EN \\
\midrule
0.20 & BOP2-F & 9.6 & 88.9 & 0.0 & 88.9 & 20.3 \\
 & BOP2-FE-1 & 10.1 & 82.6 & 9.0 & 91.6 & 20.4 \\
 & BOP2-FE-2 & 9.9 & 83.7 & 8.8 & 92.5 & 19.2 \\
 & CEVAM-NT & 10.1 & 70.1 & 10.0 & 80.0 & 23.2 \\
 & CEVAM-T & 10.1 & 83.2 & 9.4 & 92.6 & 17.1 \\
\addlinespace
0.30 & BOP2-F & 52.1 & 46.0 & 0.0 & 46.0 & 31.1 \\
 & BOP2-FE-1 & 53.6 & 37.5 & 49.1 & 86.6 & 24.9 \\
 & BOP2-FE-2 & 52.4 & 40.1 & 48.5 & 88.6 & 23.1 \\
 & CEVAM-NT & 49.4 & 23.5 & 47.8 & 71.2 & 24.7 \\
 & CEVAM-T & 47.9 & 41.9 & 44.6 & 86.5 & 19.8 \\
\addlinespace
0.40 & BOP2-F & 88.3 & 11.4 & 0.0 & 11.4 & 37.7 \\
 & BOP2-FE-1 & 89.2 & 8.9 & 86.9 & 95.8 & 21.7 \\
 & BOP2-FE-2 & 87.8 & 10.2 & 86.0 & 96.2 & 20.4 \\
 & CEVAM-NT & 87.3 & 4.2 & 85.8 & 90.0 & 18.6 \\
 & CEVAM-T & 84.2 & 12.9 & 81.8 & 94.7 & 16.6 \\
\addlinespace
0.50 & BOP2-F & 98.2 & 1.8 & 0.0 & 1.8 & 39.5 \\
 & BOP2-FE-1 & 98.4 & 1.5 & 98.1 & 99.6 & 16.7 \\
 & BOP2-FE-2 & 98.5 & 1.5 & 98.2 & 99.7 & 16.3 \\
 & CEVAM-NT & 98.8 & 0.5 & 98.5 & 99.1 & 13.3 \\
 & CEVAM-T & 97.4 & 2.4 & 97.0 & 99.4 & 13.0 \\
\bottomrule
\end{tabular}
\begin{tablenotes}[flushleft]
\footnotesize \item PRN, probability of rejecting the null; ETF, early termination for futility; ETE, early termination for efficacy; PET, probability of early termination; EN, expected sample size.
\end{tablenotes}
\end{threeparttable}
\end{table}

\begin{table}[t]
\centering
\caption{CEVAM boundaries for the TREND trial re-analysis. The null and target pCR rates were \(p_0=0.41\) and \(p_1=0.56\), respectively, with nominal one-sided type I error level \(\alpha^\ast=0.05\) and maximum sample size \(N=65\).}
\label{tab:trend-cevam-boundaries}
\begin{tabular}{lccccc}
\hline
Method & Look \(n\) & \(\alpha_{\mathrm{eff}}\) & \(\beta\) & Futility boundary & Efficacy boundary \\
\hline
CEVAM-NT & 44 & 0.13 & 0.20 & 18 & 25 \\
CEVAM-NT & 65 & 0.13 & 0.20 & 28 & 35 \\
CEVAM-T  & 44 & 0.20 & 0.82 & 21 & 24 \\
CEVAM-T  & 65 & 0.20 & 0.82 & 31 & 35 \\
\hline
\end{tabular}
\end{table}

\end{document}